\documentclass{aa}
\usepackage{graphics}
\def\laeq{\;\raise.2ex\hbox{$<$}\kern-.75em\lower.9ex\hbox{$\sim$}\;}
\def\gaeq{\;\raise.2ex\hbox{$>$}\kern-.75em\lower.9ex\hbox{$\sim$}\;}
\begin{document}
\setlength{\topmargin}{-1.0cm}
   \thesaurus{01     
              (02.13.1;  
               02.18.5;  
               03.13.2;  
               08.16.7;  
               13.07.2;  
               13.25.3)} 
   \title{Limits on magnetic field strength of 
the extended nebula of PSR~B1706-44 
         from optical, X-ray and TeV observations}

   \author{R. R. Sefako\inst{1}\fnmsep\thanks{Email:
          rocky@fskbhe3.puk.ac.za}, O. C. de Jager\inst{1}, D. J. Van der Walt
           \inst{1}
           \and
           H. Winkler\inst{2} 
          }
%
%
   \institute{Space Research Unit, Potchefstroom University, Potchefstroom,
              2520, South Africa\\
          \and
             Dept of Physics, Soweto Campus, Vista University, Bertsham
             2013, Johannesburg, South Africa\\
             }


   \titlerunning{Limits on `nebular' magnetic field strength of PSR~B1706-44}
   \authorrunning{R. R. Sefako et al.}
\maketitle
\begin{abstract}
The detection of unpulsed TeV emission from PSR~B1706-44 raised the
question if this VHE emission results from inverse Compton scattering
of multi-TeV electrons on the CMBR and other soft photon fields. 
This hints at the existence of
an unseen synchrotron nebula, which is larger than the observed 
$1'$ compact X-ray nebula. In a search for the synchrotron counterpart of
the TeV nebula, we
have taken V-band CCD images of the region around 
the PSR~B1706-44 pulsar position. By rejecting all
point source contributions down to the diffuse sky limit, we were able to
search for such extended plerionic synchrotron emission on scale
sizes limited by the angular resolution of the CANGAROO imaging TeV
observations. By combining the V-band flux limits with the observed
TeV flux, we were able to obtain upper limits for the magnetic field 
strength as a function of the radius of the assumed TeV nebula.
Assuming that the $2'$ radio plerion also defines the size of the
TeV plerion, we
constrain the steady state particle spectral index to values between
4 and 5.5, with a field strength limit of about 3~$\mu$G.
This compares with an index of $3.6\pm 0.6$ inside the 1' compact
X-ray nebula.

\end{abstract}

\section{Introduction}
PSR B1706-44 is a young Vela-like pulsar (spin-down age 
17.5~kyr), with a period of 102 ms and a large spin-down power
(\.{E}~=~3.4~x~10$^{36}$~ergs/s). It was discovered during a 20~cm 
radio pulsar 
survey of the southern Galactic plane (Johnston et al. 1992), and later
detected in soft X-rays during the ROSAT mission (Becker et al. 1992, 1995) 
and
identified as a pulsed GeV source by EGRET (Thompson et al. 1992). 
Very high
energy (VHE) $\gamma$-ray observations above 1~TeV from CANGAROO 
(Kifune et al. 1995; Kifune 1997) and above 0.3~TeV from Durham 
(Chadwick et al. 1997) detected and
confirmed the existence of unpulsed radiation of statistical significance from 
this source. The CANGAROO detection is consistent with a point source, with
angular extend not exceeding $\sim 0.^{\circ}12$ -- the pixel size of the 
imaging camera.  
A dispersion based distance measure of Taylor \& Cordes (1993)
places PSR B1706-44 at $\sim$1.8~kpc. A possible SNR G~343.1-2.3 
association 
with the pulsar was proposed by McAdam et al. (1993), but this was
later found to be unlikely by Frail et al. (1994) and 
Nicastro et al. (1996).        

Chakrabarty \& Kaspi (1998) 
gave a red-band upper limit of $R\gaeq18$ to the pulsar. A 3$\sigma$ upper
limit to the pulsar magnitude of $V=24.5$ was given by Lundqvist et al.
(1999), and this is consistent with the theoretical prediction of $V=24.12$
(Urama \& Okeke 1998, and references therein). Mignani et al. (1999), using
the same data as Lundqvist et al., obtained an upper limit of $V\gaeq 27.5$. 

The photon flux above 1~TeV from this unpulsed source is
only two times smaller compared to the flux of the Crab Nebula at the
same energy. This is remarkable in the context of a synchrotron-inverse 
Compton (on external photon fields) 
interpretation for the TeV $\gamma$-rays from PSR B1706-44, since
the synchrotron nebula of this source is very weak compared to the Crab
Nebula's synchrotron intensity. Whereas the Crab Nebula's intense
synchrotron emission is the result of a large $B$ ($\sim 10^{-4}$~G)
field, we must have a much weaker field for this plerion to avoid 
a bright synchrotron nebula. 

De Jager (1995) speculated that a Vela-like compact synchrotron
nebula ($<1'$, Harnden et al. 1985) may be present, which would
account for most of the unpulsed X-ray emission from this source,
since the pulsar wind magnetic field (which scales as $B\propto 1/R$)
would predict a compact synchrotron nebula, if the conditions are similar
to Vela. The corresponding 
particle density required for the synchrotron nebula
would however be too low to produce a detectable 
inverse Compton compact nebula, since the target photon density from the
CMBR and Galactic disk would be too small.

A compact nebula was indeed discovered
(Finley et al. 1998), which confirms the abovementioned interpretation.
The scenario of Harding \& de Jager (1997), with a detailed analysis
by Aharonian et al. (1997) may apply in this case:
electrons are streaming away
from the compact X-ray nebula into a low-B extended plerion.
If the diffusion coefficient for electrons in the extended plerion is
small enough, it may be possible to trap enough electrons to account for
the TeV emission by IC scattering.

Frail et al. (1994) detected a radio synchrotron nebula associated
with the pulsar, which has a radius of $\sim 2'$ at 20~cm wavelength.
This may represent the low energy counterpart of the TeV nebula, since
the associated $B$ in the radio nebula is smaller than the pulsar driven 
field in the compact X-ray nebula, and the size is just smaller than the
angular resolution of the CANGAROO telescope. 

If the TeV $\gamma$-rays originate from the extended (radio?) nebula,
we will expect a bright extended nebula in optical and/or X-rays,
if the field strength is large enough. No extended X-ray nebula was 
however seen, resulting in an upper limit of four times the flux of the point 
source/compact X-ray nebula at 1~keV, for $R=2'$ (K. Brazier 1998, personal 
communication).
This is consistent with the expectation that the pulsar field
strength (which drops as $1/R$ outside the light cylinder) should start
to drop below the ambient field strength of a few $\mu$G at a distance
of a few arcminutes (de Jager \& Harding 1998).
The consequence of this is that both the synchrotron brightness
and synchrotron characteristic frequencies should drop with increasing
$R$. Scaling from the Durham detection above 0.3~TeV, and the 
expected $B\sim3\;\mu$G in the extended nebula (de Jager \& Harding 1998),
we may use equation (5) of Aharonian et al. (1997) to calculate the 
synchrotron frequency
which corresponds to the Durham detected $\gamma$-rays near 0.3~TeV,
assuming that the CMBR is mostly responsible for the inverse Compton
scattering. This gives a frequency of
$$\nu\sim 1.5\times 10^{15} \left(\frac{E_{\gamma}}{0.3\; TeV}\right)
\left(\frac{B}{3\;\mu G}\right)\; Hz,$$
which is within a factor of 3 from the V-band. 
The non-detection of the unpulsed component at 20 GeV (Thompson et al. 1996) 
hints at a marginal turnover in the $\gamma$-ray spectrum (de Jager \&
Harding 1998),
which corresponds to a turnover frequency as low as $10^{14}$ Hz, which
is well below the V-band. 

At a distance of 1.8 kpc to the source, we expect significant interstellar
absorption
in the blue and UV bands. We therefore take wide field V-band images
as a compromise between interstellar absorption and avoiding
a spectral turnover at too low frequencies. By taking overlapping images,
we cover a total field-of-view of about $10'$, which 
should include the expected TeV source. Even if the full extend of the TeV
source is $10'$, we still expect a typically centrally brightened
image, resulting in a radial gradient from the source. 
Failure to
detect the optical counterpart will allow us to set upper limits
on the magnetic field strength in the extended nebula, since 
any stronger field should have resulted in a brighter optical nebula.

\section{Observations and data analyses}

{\it V} band CCD observations of PSR~B1706-44 were carried out 
on May 24, 1999 at
the South African Astronomical Observatory (SAAO) 1.0~m telescope in 
Sutherland. Five
frames, each $\sim5.^{'}3\times 5.^{'}3$ (pixel size = 0.31 arcsec), 
were taken. The first one was
centered on the radio pulsar position (Johnston et al. 1992, 1995), 
whereas
the other four frames were offset in such a way that they have an overlap of
$\sim$10\% with each other around the pulsar position. The
overall region observed around the pulsar is 
$9.^{'}5\times9.^{'}5$. The observation time for each frame was 12 minutes.
Table~\ref{tab:obs} gives the position (RA, Dec), the starting time (MJD),
$K$ values and the duration time (seconds) log for the five frames. 

\begin{figure}
\resizebox{\hsize}{!}{\includegraphics{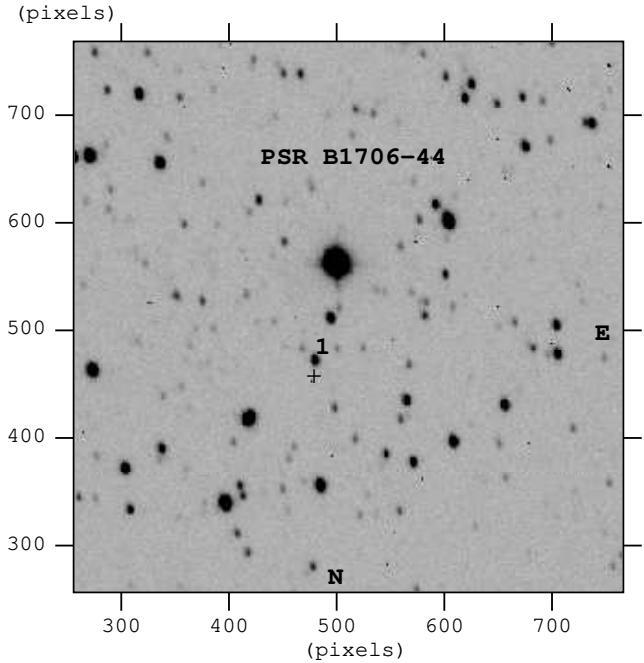}}
\caption{Star {\bf 1} is a standard star, {\bf +} is a pulsar 
               position (Mignani et al. 1999). The frame is zoomed
               in. The original frame is about 1024 x 1024, CCD frame size.}
\label{fig1}
\end{figure}

Observations were affected by the moonlight (about 60\% bright). The average 
counts
for each frame were decreasing with time, probably due to the moon's effect
across the sky. The data were cleaned and flat-fielded using
the IRAF image-processing software. Fig.~\ref{fig1} shows a {\it V}
band CCD image of the
central frame with the pulsar and the standard star positions labeled {\bf
+} and {\bf 1} (Mignani et al. 1999, their Figs. 1 \& 2), 
respectively. Only
the inner part (25\% of the total area) of Frame 1 is shown on 
Fig.~\ref{fig1}.   

\begin{table}
\caption{Observing log for B1706--44. The 1st frame is
the central one.}
\label{tab:obs}
\begin{tabular}{c c c c c} 
\hline
Frame& RA (J2000) Dec& MJD (start)& K& time(sec)\\ 
\hline
1 & 17 09 42 -44 28 57 & 51322.80399&29.7&720\\
2 & 17 09 52 -44 31 20 & 51322.81410&28.7&720\\
3 & 17 09 33 -44 26 34 & 51322.82833&29.2&720\\
4 & 17 09 52 -44 26 34 & 51322.83872&28.3&720\\
5 & 17 09 33 -44 31 20 & 51322.85028&28.6&720\\
\hline
\end{tabular}
\vspace{4mm}
\end{table}

\begin{figure}
\resizebox{\hsize}{!}{\includegraphics{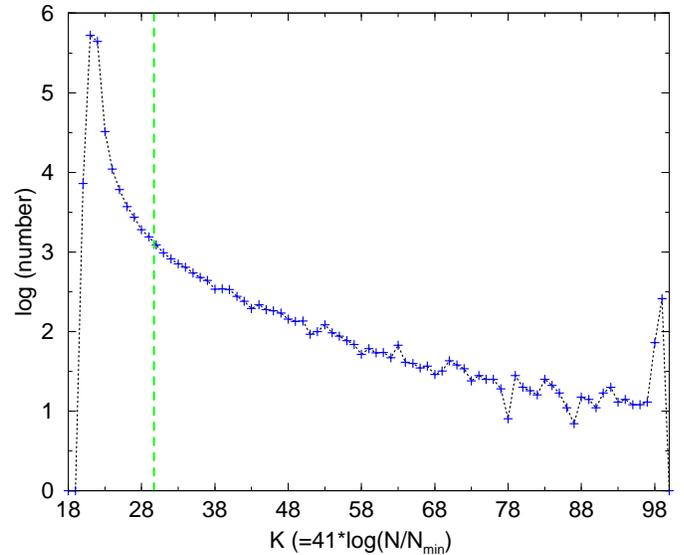}}
\label{fig2}
\caption{A non-linear intensity distribution histogram of the central 
frame (example). $K=1$ corresponds to 264 counts (minimum), whereas $K=100$
corresponds to $6.89\times10^{4}$ counts (maximum). The cut is 
made at $K\sim30$, above which the 
high intensity star light contribution is rejected.}
\end{figure}

For each frame, we made a histogram of the number of pixels versus
intensity per pixel using a logarithmic binning procedure prescribed by 
$$K = N_{o}\log (N/N_{min}),$$  
where $N_{o}$ ($= 99/\log (N_{max}/N_{min})$) is a normalizing 
constant, as  shown in Fig.~\ref{fig2} for the central frame. This binning 
procedure ensures that all intensities are confined to the interval $K=1$ and
100. 

Whereas the diffuse emission is expected to be confined to a rather narrow
low-$K$
peak in the histogram, the pixels contaminated by point sources should show
a broad tail. Fig.~\ref{fig2} indeed shows such behaviour. It can be
shown that the weak plerionic emission is expected to contribute to the low
$K$-values ($K\laeq 25$), so that we can reject the star--like objects by
cutting at $K\geq 30$ (for frame~1) as shown in Fig.~\ref{fig2}. This
rejects only 10\% of the pixels, so that we preserve 90\% of the plerionic 
component.  The threshold value of K for
each frame is given in Table~\ref{tab:obs}, and probably depends on the 
phase of the moon and the background bright stars (in each frame), which
contribute to the background for each frame. Note that the final results 
are not sensitive to the threshold $K$.  

By renormalising the count rate of each frame to the count rate of the
central frame, we were able to construct a composite image of a
$\sim 10^{'}\times 10^{'}$ region around the pulsar. Since the plerionic
structure is expected to be circular around the pulsar (the radio nebula is
at least circular), we constructed annuli around the pulsar and made a plot
of the mean count rate per pixel
versus radius of the annuli. These count rates
are constant with radius within statistics for $R=0$ to $5^{'}$,
which means that there is no evidence for a centrally brightened source, and
hence no evidence for a plerionic structure centered on the pulsar.

\section{Upper limits} 
\begin{figure}
\resizebox{\hsize}{!}{\includegraphics{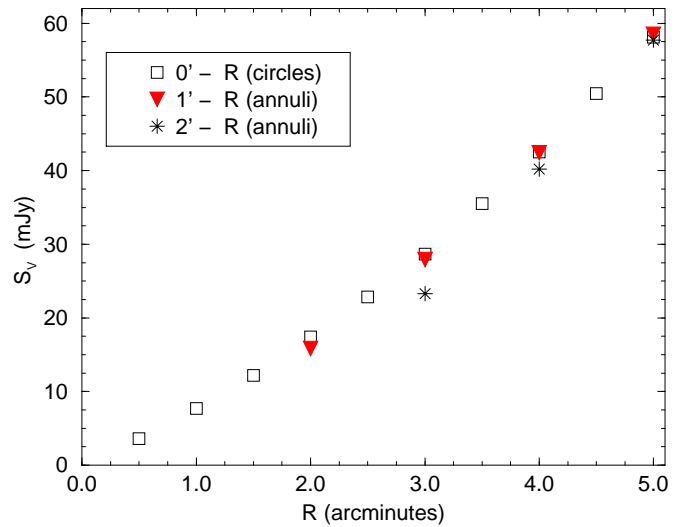}}
\caption{The $3\sigma$ upper limit to the optical flux, $S_{V}$ (mJy), vs 
radius, R (arcminutes) from 
the radio pulsar position. The `filled triangles' and the 
`stars' represent 
annuli from 1 and 2 arcminutes, respectively. The `open squares' represent the
enclosed region inside each circle with radius R.}
\label{fig3}
\end{figure}

The 3$\sigma$ upper limits for
plerionic emission confined in circles, as well as annuli, are shown in
Fig.~\ref{fig3}. These upper limits were converted to absolute intensities,
$S_{V}$, using the observed counts from Star {\bf 1} (see Fig. \ref{fig1}) 
of Chakrabarty \& Kaspi (1998) with $V=17.4$~mag. 

The 3$\sigma$ $S_{V}$ upper limits increase with $R$ as expected, and at
$R=0.^{'}5$ (compact X-ray nebula), our 3$\sigma$ upper limit is $V\geq15$,
while for an extended nebula of about 4~arcminutes, the upper limit is
$V\gaeq 12$.  

\begin{figure}
\resizebox{\hsize}{!}{\includegraphics{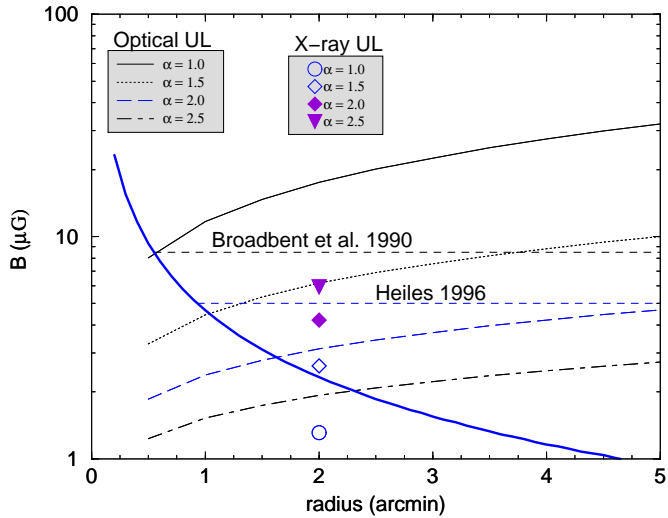}}
\caption{Plots of magnetic field, B ($\mu$G), against radial
distance, R (arcminutes) from the pulsar. 
The thick solid line (lower limit) represents the pulsar driven
field strength, assuming a $\sigma=1$ unshocked wind 
(Kennel \& Coroniti 1984) at large radii. For comparison, 
the Broadbent et al. (1990) 
and Heiles (1996) Galactic magnetic field strength values are also given.}
\label{fig4}
\end{figure}

Assuming that the synchrotron spectral index, $\alpha$, of the nebula 
is the same in 
both the optical and X-ray regions, the monochromatic  X-ray flux, 
$S_{1keV}$,  at 1~keV is given by
\(
S_{1keV}=S_{V}({\epsilon}_{x}/{\epsilon}_{o})^{-\alpha},
\)
where ${\epsilon}_{x}$ (=~1~keV) and ${\epsilon}_{o}$~(=~2.3~eV) are 
the X-ray and optical energies, respectively.
The observed optical flux can be expressed as $S(V)=S_{V}\exp(-\tau_{\nu})$,
where $\tau_{\nu}$ is the optical depth
due to interstellar absorption, with $\tau_{\nu}\sim 1.8$ for $d=1.8$~kpc. 
Using the above relations and the de 
Jager et al. (1995) expression for the TeV $\gamma$-ray IC energy spectrum
(their Equation 2), we calculate the upper limits to the magnetic fields 
for different values of $\alpha$ between 1.0 and 2.5, inclusively.
Fig.~\ref{fig4} shows the curves for $\alpha$ = 1.0, 1.5, 2.0 and 2.5, as
well as the magnetic field limits from the X-ray upper limits for 
$R=2'$ for similar values of $\alpha$. The
magnetohydrodynamic curve of Kennel \& Coroniti (1984) model, with 
${\sigma}=1.0$ (equipartition between particles and fields), is 
superimposed on Fig.~\ref{fig4}, and it is considered as
the lower limit to the extended nebula at large radii ($R\gaeq 2.^{'}0$).
Also shown in Fig.~\ref{fig4} are the Galactic field strengths 
at the source according to Heiles (1996) and Broadbent et al. (1990).

\section{Discussions}
Our search for an extended nebula synchrotron counterpart to the 
TeV nebula did not reveal any nebular structure around PSR~B1706-44. 
The TeV observations do hint at the existence of such a plerionic
structure, but our results constrain the nebular magnetic field
strength if we assume that the TeV $\gamma$-rays originate from the
IC scattering of energetic pulsar wind electrons
on the CMBR. At $R=2'$, we find that the optical and X-ray limits are
complementary: the spectral index is constrained to $\alpha = 1.5$ to 2.3,
whereas $B\leq 3$~$\mu$G. This limit is barely above the pulsar
wind solution, but below
the average Galactic field values shown in Figure 4. Furthermore,
our constraints on $\alpha$ are marginally consistent
with $\alpha = 1.3\pm 0.3$ measured for the compact X-ray nebula 
(Finley et al.  1998).
 
Our optical observations and the information gathered from other wave bands
(mainly from X-ray and $\gamma$-ray energies) show that 
better quality observations (dark moon) should improve these limits, or,
result in a detection. Polarimetry of the 
``extended nebula" of PSR~B1706-44 may 
also help to clarify the status of synchrotron emission.




\end{document}